\documentclass[aps,prl,twocolumn,superscriptaddress]{revtex4-1}
\usepackage{longtable}
\usepackage{amsfonts}
\usepackage{amsmath}
\usepackage{graphicx}
\usepackage{stackengine,graphicx}
\usepackage{epstopdf}
\usepackage[caption=false]{subfig}
\usepackage[colorlinks]{hyperref}

\begin{document}

 \title{Localization of Trivial Edge States in InAs/GaSb Composite Quantum Wells}

\author{Vahid Sazgari}
\affiliation{Faculty of Engineering and Natural Sciences, Sabanci University, Tuzla, 34956 Istanbul, Turkey}
\affiliation{Sabanci University Nanotechnology Research and Application Center, Tuzla, 34956 Istanbul, Turkey}
\author{Gerard Sullivan}
\affiliation{Teledyne Scientific and Imaging, Thousand Oaks, CA 91630, USA}
\author{\.{I}smet \.{I}. Kaya}
\affiliation{Faculty of Engineering and Natural Sciences, Sabanci University, Tuzla, 34956 Istanbul, Turkey}
\affiliation{Sabanci University Nanotechnology Research and Application Center, Tuzla, 34956 Istanbul, Turkey}

 \email{iikaya@sabanciuniv.edu}

 \begin{abstract}

InAs/GaSb heterostructure is one of the systems where quantum spin Hall effect is predicted to arise. However, as confirmed by recent experimental studies, the most significant highlight of the effect i.e., the  conductance quantization due to non-trivial edge states is obscured by spurious conductivity arising from trivial edge states. In this work, we present experimental observation of strong localization of trivial edge modes in an InAs/GaSb heterostructure which was weakly disordered by silicon delta-like dopants within the InAs layer. The edge conduction which is characterized by a temperature-independent behavior at low temperatures and a power law at high temperatures is observed to be exponentially scaled with the length of the edge. Comprehensive analysis on measurements with a range of devices is in agreement with the localization theories in quasi one-dimensional electronic systems. 
\end{abstract}

\date{\today}
\maketitle
Topological Insulator (TI) is a new phase of matter that has attracted a great interest not only for fundamental scientific research but also for its potential applications. Specifically, the quantum spin Hall (QSH) insulator is a two-dimensional topologically non-trivial insulator that is theoretically predicted to manifest an insulating bulk but accompanied by dissipationless helical states at the edges~\cite{Kane_PRL2005,BHZ_Science}. These new class of materials provide an ideal platform for novel spintronic applications, quantum information, and quantum engine~\cite{Spintronics1,Spintronics2,Inanc}. The QSH effect was first predicted and soon observed in band inverted HgTe/(Hg,Cd)Te quantum wells (QWs)~\cite{BHZ_Science,Konig_Sci2007,Roth_Sci2009}, where a transition from normal to topological phase can only be tuned by the thickness of the HgTe QW. Among the new proposals for a QSH system,
the InAs/GaSb bilayer QW heterostructure is favored as it benefits from high mobility, easy fabrication, and most importantly the tunability of the band structure
by an electric field~\cite{Liu2008,KnezPRL107,SuzukiPRB2013,NichelePRL2014,KnezPRL2014,SpantonPRL2014,DuPRL2015,QuPRL2015,NicheleNJP18,NguyenPRL117,ShojaeiPRM2}.
Particularly, by application of dual gates on top and bottom of the heterostructure, the Fermi level and the alignment between InAs conduction and GaSb valence bands can be controlled independently in such a way that a continuous transition from trivial to broken-gap nontrivial insulator is possible~\cite{Liu2008,QuPRL2015}.

In the early experiments on InAs/GaSb double QWs, the edge conduction was overshadowed by relatively large residual bulk conductance~\cite{KnezPRL107}. Subsequent efforts to suppress the bulk conductance via localization of bulk carriers incorporated the introduction of Si impurities at the interface between InAs and GaSb quantum wells~\cite{KnezPRL2014,SpantonPRL2014,DuPRL2015}, Be doping in the QW barrier~\cite{SuzukiPRB91,CouedoPRB94}, or using low purity Ga source material for the GaSb layer~\cite{CharpentierAPL103,MuellerPRB92}. These studies confirmed robust edge conduction, however whether the conduction was due to helical or trivial edge states was not proven. On the other hand, more recent works demonstrated the presence of edge transport also in the trivial phase of these heterostructures~\cite{NicheleNJP18,NguyenPRL117,MuellerPRB96}. The experiments with different sample geometries and edge lengths concluded that the edge transport is nearly identical in both trivial and inverted phases of their heterostructures, where the edge resistance consistently increased with the edge length in a linear fashion. Measured resistance plateaus situated well below $h/e^2$ for small samples evidenced for multi-mode non-helical edge states. This trivial edge conductance has been attributed to band bending at the vacuum interface, fabrication-induced spurious effects, or concentrated electric field at the edges of the sample induced by the top gate covering the etched steps~\cite{NicheleNJP18}. Coexistence of non-helical and helical edge channels in InAs/GaSb bilayers in the inverted regime obscures the QSH effect and demands further experimental efforts for the elimination of trivial edge conductance.

While the helical edge states benefit a topological protection against perturbations which are symmetric under time reversal, the trivial states are naturally sensitive to such perturbations. According to localization theories, energy states in a disordered electronic system are localized at low temperatures provided that the system size is much larger than the electron mean free path~\cite{Loc1,Loc2,Loc3,Loc4}. As a result, as the temperature is reduced, a trivial 1D conductor tends to an insulating state. On the other hand, for a 1D spin-momentum-locked edge state in QSH phase, elastic backscattering is prohibited due to the energy cost of spin to flip, and only a process breaking the time reversal symmetry or an inelastic scattering allowing the spin-flip may destroy the quantization of edge conductance. Several backscattering mechanisms that have been proposed include magnetic impurities~\cite{MagScat1,MagScat2,MagScat3,MagScat4,MagScat5,MagScat6}, magnetic-field-induced localization~\cite{MagScat7}, nuclear-spin-induced backscattering via hyperfine interaction~\cite{NucScat1,NucScat2,NucScat3,NucScat4}, inelastic interactions induced by nonuniform Rashba spin orbit coupling~\cite{Rashba1,Rashba2,Rashba3}, phonons~\cite{PhonScat}, multiparticle scattering~\cite{Multi1,Multi2,Multi3} and charge puddles~\cite{CharPuddle}.    
 
Future experiments look for a non-invasive time-reversal-invariant perturbation to eliminate the trivial edge states whereas the topological phase remains intrinsically immune. One way to suppress the edge conduction in the trivial phase is to achieve the localization of edge carriers by introduction of adequate level of impurity. A weak disorder may not be able to fully localize the one dimensional edge carriers, whereas a strong disorder may induce sufficiently large potential fluctuations which could effectively overcast the small bulk gap in the inverted phase and lead to a semi-metallic behavior. Based on this expectation, we deliberately disordered our heterostructures by Si delta-doping to investigate its effect on the trivial edge conduction.

Here, we report the localization of the trivial edge states in an InAs/GaSb heterostructure with Si delta-doping near the interface of the two quantum wells inside the InAs layer. While the heterostructure retains its high mobility in the charge populated state, the Si dopants effectively localize the edge states when the system is driven into the normal insulating regime by opening an energy gap and adjusting the Fermi level within it, using the dual gates on top and bottom. The edge dominant transport is verified by non-local measurements in a Hall bar device and temperature-dependent conductance measurements in a Corbino disk. Temperature dependence of the edge conductance within the insulating phase in the Hall bar shows the localization behavior in a quasi-one-dimensional disordered conductor. All together, we confirmed the localization of the trivial edge states as it will be discussed in the following paragraphs.

The heterostructure was grown on a GaAs substrate by molecular beam epitaxy. Following a  $1~\mu$m AlGaSb buffer layer, the bilayer QW structure was grown, which consists of 12~nm InAs on 9~nm GaSb sandwiched between top and bottom AlGaSb barriers of 50~nm. Silicon atoms, as a single layer impurity, were added in the form of delta doping during the growth of InAs layers. The Si particles with an average density of $10^{11}$ cm$^{-2}$ were deposited inside the InAs at 2 atomic monolayer distance from the InAs/GaSb interface. The top barrier was protected from oxidization by a 3~nm GaSb cap layer. A 20$\times$5~$\mu$m ($L \times W$) Hall bar device with uniform gate, a 150$\times$10 $\mu$m Hall bar with finger gates and a Corbino disk which is top-gated by a ring-shaped metal with inner and outer diameters of $r_i$=400~$\mu$m and $r_o$=600~$\mu$m respectively were used in the measurements. The device mesas were patterned by standard electron beam lithography and defined via chemical wet etching. The etched surfaces were passivated by a 100~nm PECVD-grown Si$_{3}$N$_{4}$ layer which also served as the dielectric  separating the Ti/Au top gate electrodes from the heterostructure. The ohmics were made by Ge/Au/Ni metalization without annealing. Transport measurements were performed in a dilution refrigerator with a base temperature of 10~mK using standard lock-in methods with 10~nA current excitation at 11~Hz, unless otherwise stated.

\begin{figure}[t]
\begin{center}
	\hspace*{1.5mm}
\includegraphics[width=0.48\textwidth]{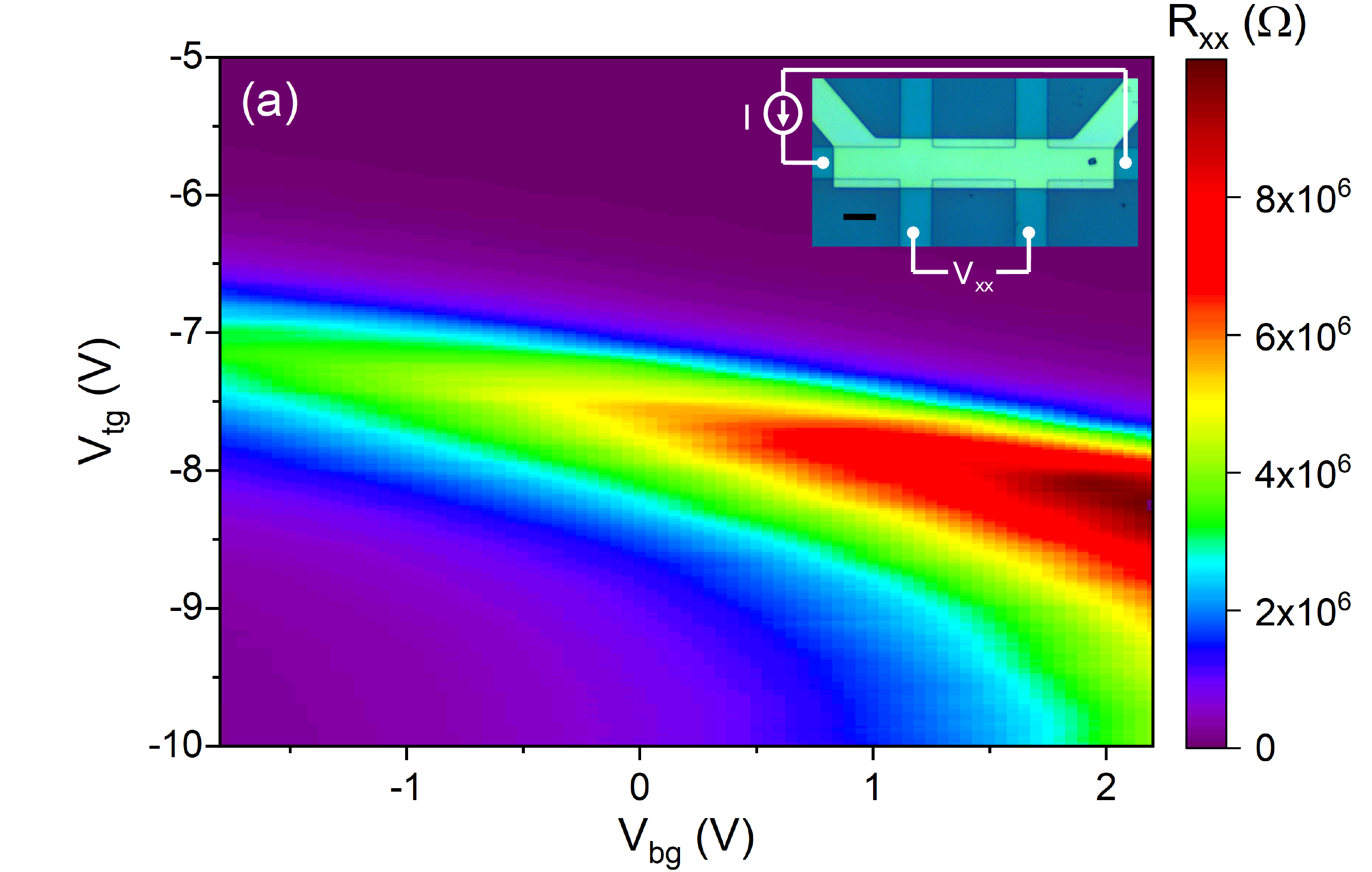}
\includegraphics[width=0.48\textwidth]{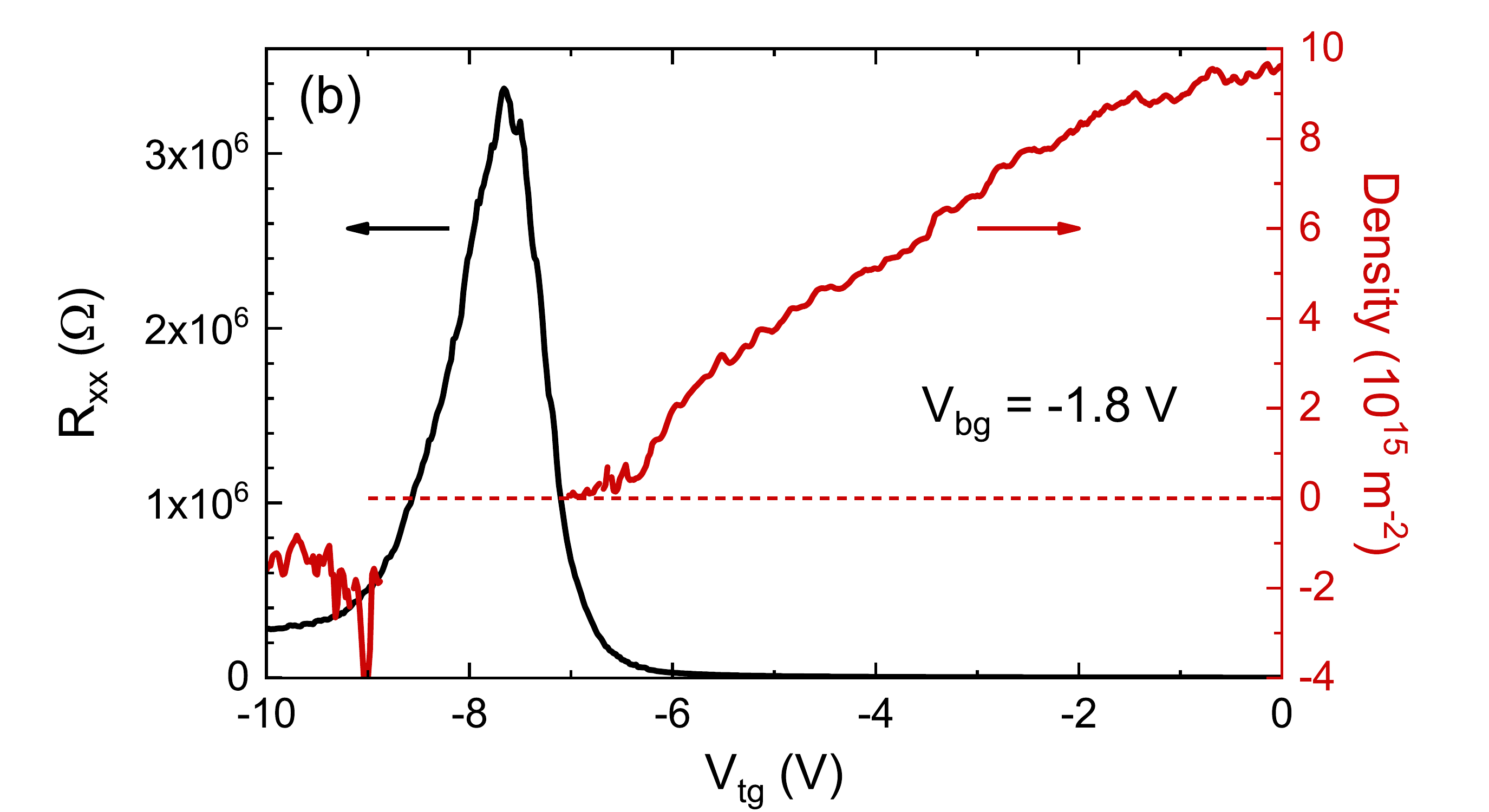}
\includegraphics[width=0.48\textwidth]{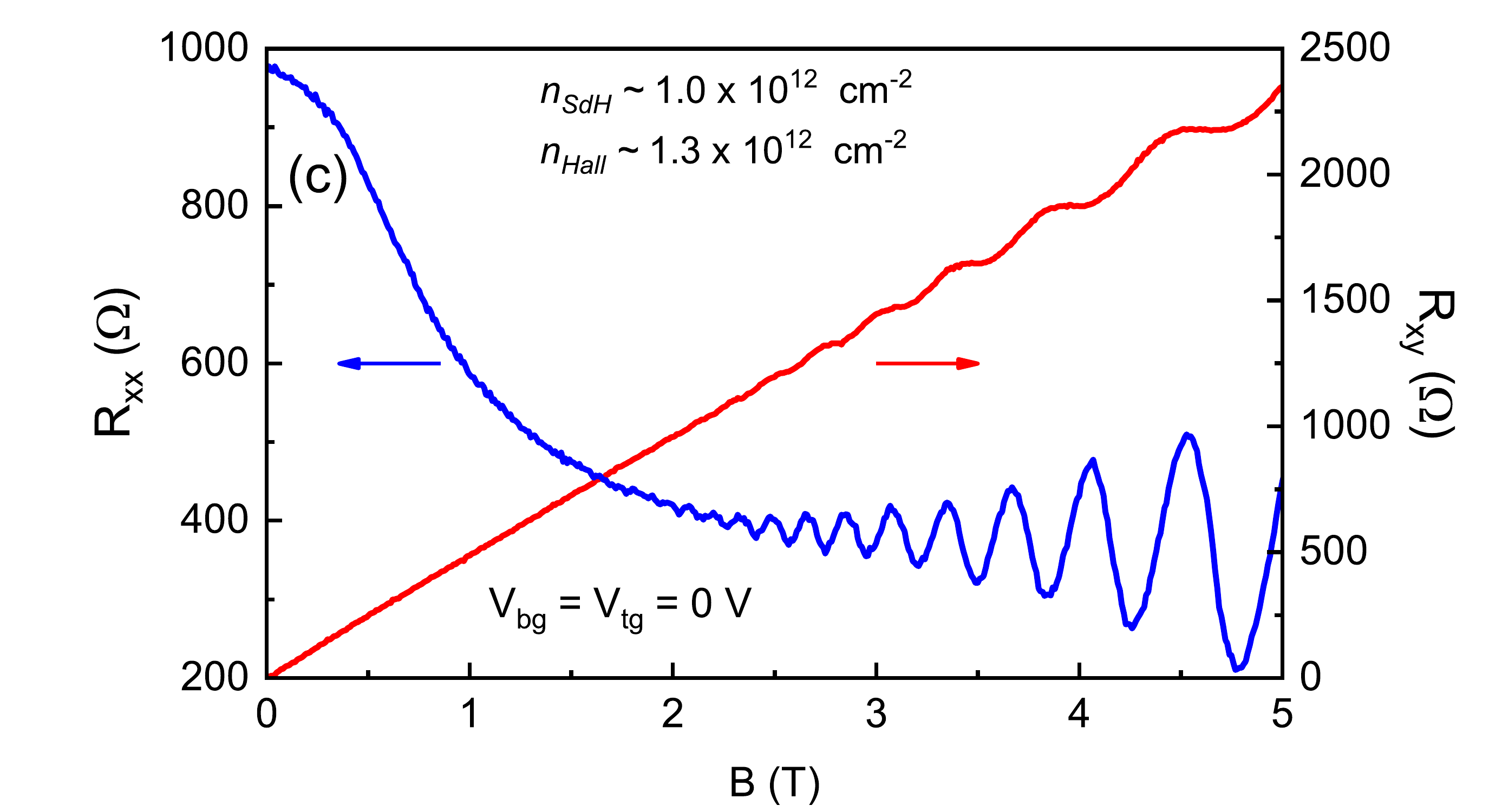}
\caption{\label{Fig1} (a) Longitudinal resistance of the Hall bar device as a function of $V_{tg}$ and $V_{bg}$ measured at $T=10$~mK applying 10~nA DC current. Inset shows the optical microscope image of the device with a 5-$\mu m$ black scale bar. (b) Density modulation via top gate obtained from Hall measurements at 0.5~T (red curve). The dashed guideline indicates the zero density. The black curve shows the corresponding longitudinal resistance at B=0. (c) Longitudinal and transverse resistances as a function of B measured at 10~mK. As a signature of high mobility, clear SdH oscillations and quantum Hall plateaus are observed. The carrier densities calculated from quantum Hall and SdH oscillations are consistent with the one obtained from classical Hall effect in (b).}
\end{center}
\end{figure}

Figure~1(a) represents the longitudinal resistance of the Hall bar sample as a function of top gate, $V_{tg}$ and bottom gate, $V_{bg}$ voltages. Within the extent of the applicable gate voltages, the sample remains in the normal phase where the trivial gap decreases with decreasing $V_{bg}$. The tunability of the system was restricted within the trivial regime due to current leakage through the bottom gate for $V_{bg} < -1.8$~V.  The longitudinal resistance in the electron-rich region is about  1~k$\Omega$ and sharply increases by four orders of magnitude and the system becomes insulating when the Fermi level is tuned inside the normal gap. Thereafter, the resistance decreases as the top gate voltage is decreased and the system is moved into the hole regime, nevertheless remaining significantly higher than that of the electron side due to lower mobility and charge density. The Hall resistance measurements performed at $B=0.5$~T were used to extract the carrier density at different gate voltages. The modulation of longitudinal resistance and the carrier density by the top gate is shown in Fig.~1(b) for $V_{bg}=-1.8$~V. As $V_{tg}$ is swept towards negative voltages, the electron density continuously decreases to zero until the Fermi level lies inside the bulk gap where the longitudinal resistance peaks around 4~M$\Omega$. By further decreasing $V_{tg}$, the hole carriers are populated, though with low concentration, but below $V_{tg}=-9$~V, the carrier density slightly decreases. We attribute this to the electron release and capture in the trap states induced at the interface of GaSb and insulator layer, or the hole states induced in the GaSb cap layer, which screen the electric field from the top gate voltage~\cite{GateHysteresis,GateHysteresis2}.

Furthermore, we performed quantum Hall measurements in the Hall bar device. As it is shown in Fig.~1(c), in the electron regime, Shubnikov-de Haas (SdH) oscillations in the longitudinal resistance and well-established quantum Hall plateaus indicate a high mobility 2D electron gas in the InAs QW. The electron mobility of the sample is calculated as $\mu =$ 1.5 $\times 10^{4} $ cm$^2/$Vs at the density of $n =$ 1.0 $\times 10^{12}$~cm$^{-2}$. Assuming this mobility, the elastic mean free path is obtained as $l_e = 30~nm$ which is perfectly consistent with the mean spacing between single Si scatterers recalling that the 2D density of the silicon impurity is $10^{11}$~cm$^{-2}$. Therefore, we conclude that the silicon atoms are the dominant scattering centers leading to localization of the carriers hence an insulating behavior.  

\begin{figure}[t]
\begin{center}
\includegraphics[width=0.5\textwidth]{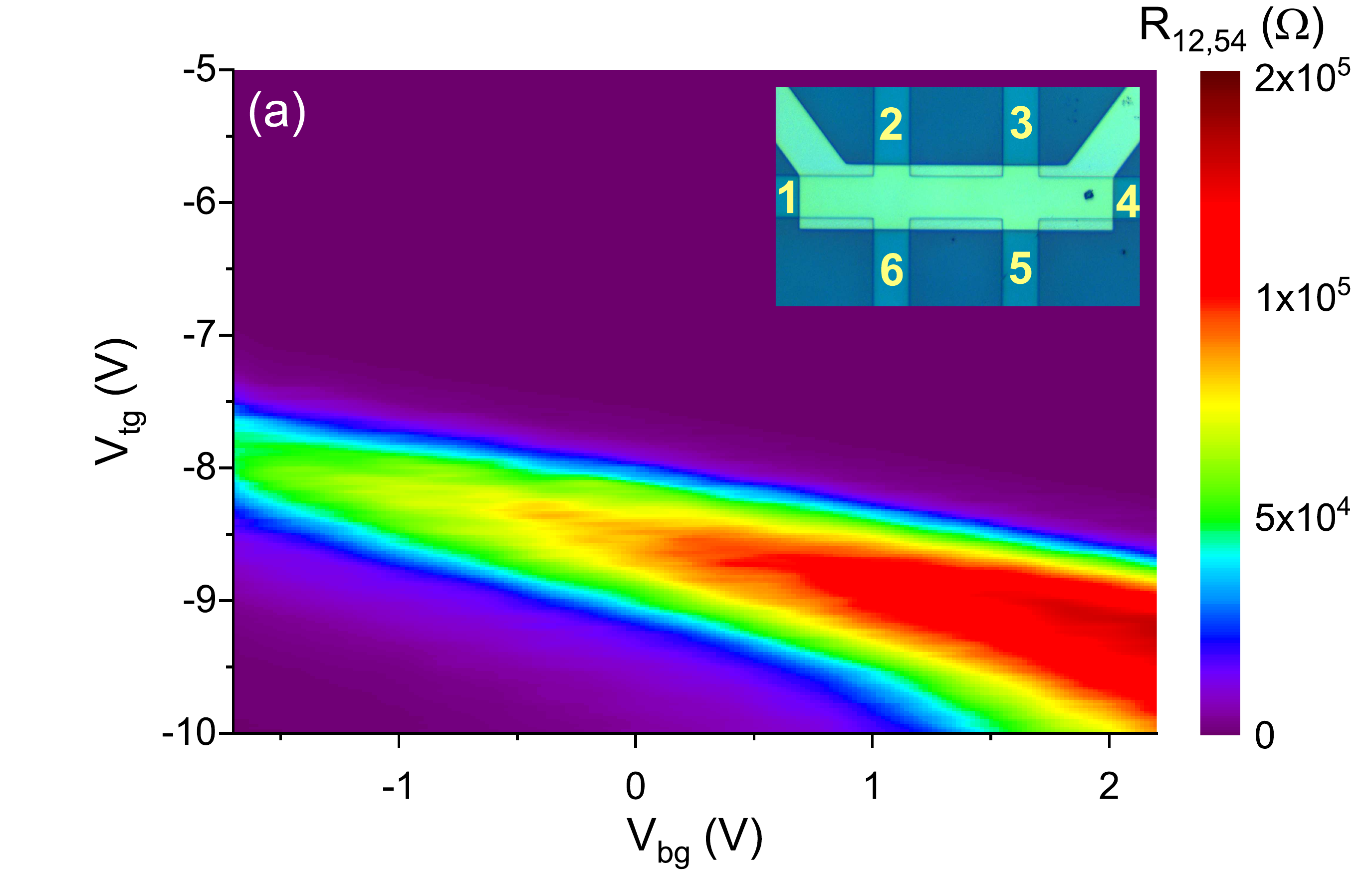}
\includegraphics[width=0.5\textwidth]{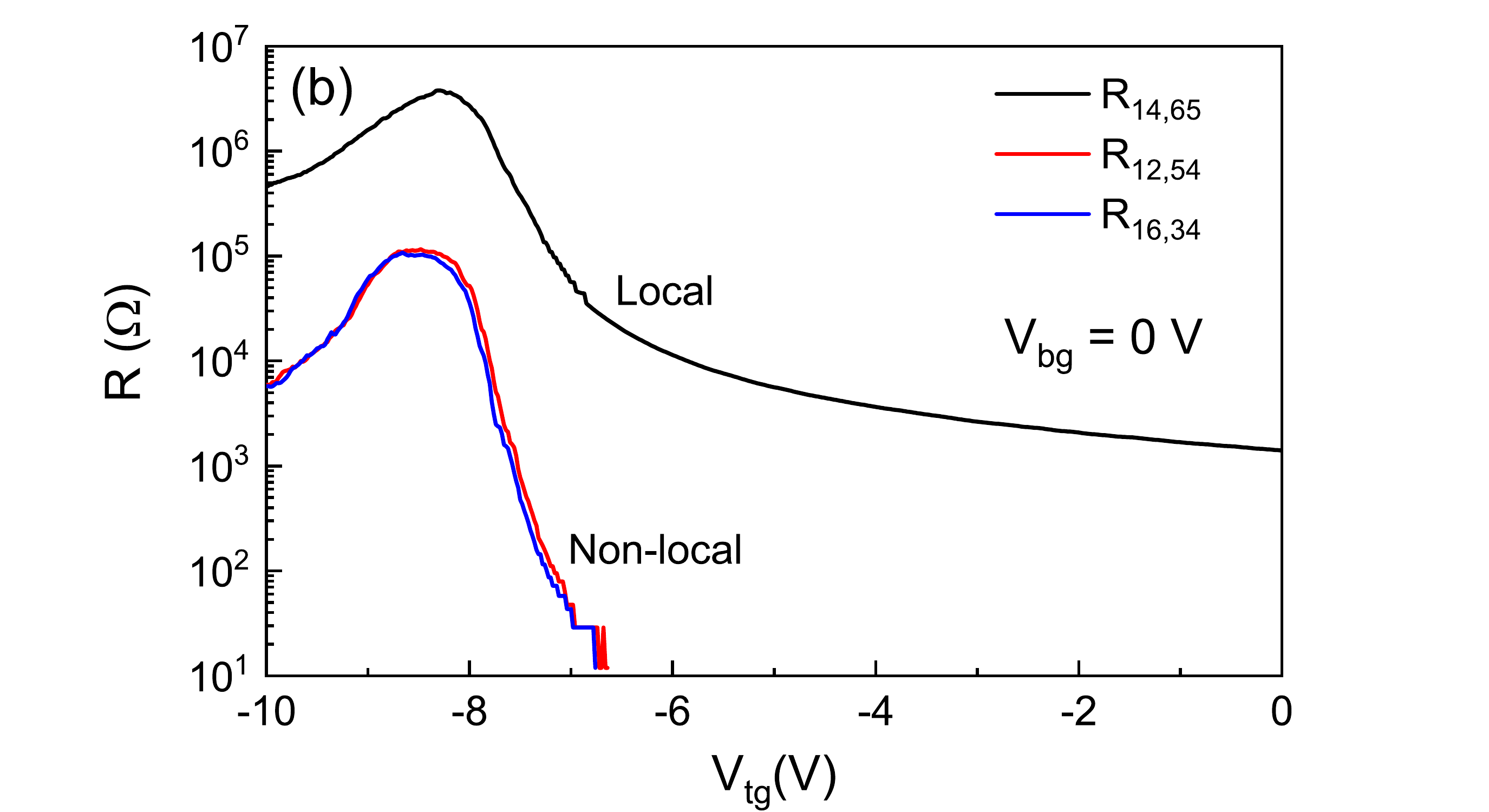}
\caption{\label{Fig2} (a) Non-local resistance measured in the Hall-bar device as a function of top and bottom gate voltages at 10~mK. Inset shows the optical image of the device with labeled contacts. R$_{ij,kl}$ is the resistance measured between k and l when the current applied between i and j. (b) Comparison of local and two different configuration of non-local resistances as functions of top gate voltage.}
\end{center}
\end{figure}

\begin{figure}[t]
	\begin{center}
		\includegraphics[width=0.5\textwidth]{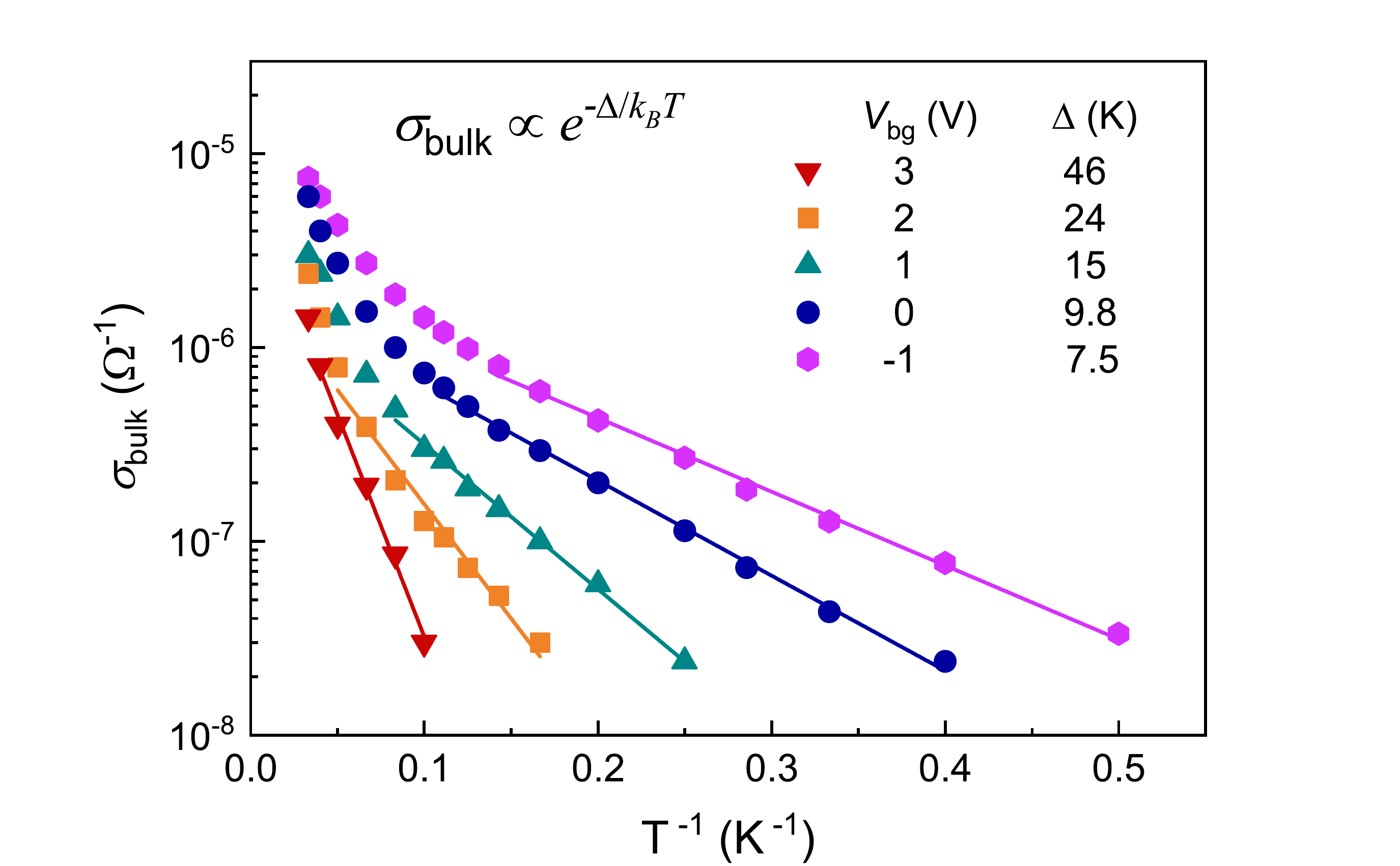}
		\caption{\label{Fig3} Temperature behavior of the minimum bulk conductivity measured in the Corbino disk with 100~$\mu$V AC bias for different bottom gate voltages. Solid lines are the fits to the Arrhenius equation giving the corresponding bulk gap for each bottom gate voltage.}
	\end{center}
\end{figure}

The Hall bar device exhibits a robust non-local signal as displayed in Fig.~2(a), which manifests conduction by edge states. As discussed below, comparison of the local and non-local signals together with the results from the Corbino device confirm that the conduction in the Hall bar within the gap is predominantly by the edge states. Fig.~2(b) compares the local and non-local resistances, with bottom gate at 0~V, as the top gate voltage is swept from electron-rich regime to the normal gap and further to the hole band. Deep in the electron regime, the non-local resistance becomes less than the measurement sensitivity. The ratio between local and non-local resistances is about 3 orders of magnitude in the measurable range in the electron side. Inside the gap, on the other hand, since the bulk conductance is suppressed while the edge is conducting, the non-local signal is magnified to within an order of magnitude of the local signal. 

In order to quantify the suppression of bulk conductance in the Hall bar device, we investigated a Corbino geometry in which the current flows radially between inner and outer contacts entirely in the bulk. The bulk conductivity of the Corbino disk is measured as a function of top and bottom gate voltages at different temperatures. The minimum conductivity in top gate sweep, which is referred to as the bulk conductivity, is calculated by $\sigma_{bulk} = (I/2\pi V) ln(r_i/r_o)$ for the ring geometry. $\sigma_{bulk}$ scales exponentially with the inverse of temperature as can be seen in Fig.~3. We calculated the bulk gap, $\Delta$ from the Arrhenius law, $\sigma_{bulk} \propto$ exp$ (-\Delta/2k_BT)$ where the fitting is excellent at temperatures $k_BT \le \Delta$. The trivial gap monotonously decreases from 46~K to 7.5~K as the bottom gate voltage is decreased from 3~V to -1~V. 

In order to examine the localization of trivial edge modes, we extracted the edge component of the Hall bar conductance. Figure~4 illustrates $G_{edge}$ of the Hall bar inside the trivial gap as a function of temperature in the range of 0.2 to 30~K for different bottom gate voltages. $G_{edge}$ is obtained by subtracting the bulk conductance, $G_{bulk} = (W/L)\sigma_{bulk}$, from the total conductance of the Hall bar. The edge conductance displays a distinct temperature behavior, saturates at low temperatures and scales with a power law at higher temperatures. Moreover, a power law with voltage bias is also observed at lowest temperature (inset of Fig.4) in accordance with the theory disscussed in the next paragraph. Here we should note that the edge conductance at low temperatures, where it saturates, is at least two orders of magnitude larger than the bulk conductance. As temperature increases, the bulk and edge conductances increase and become comparable at highest temperatures. These observations verify the dominance of edge conduction in the Hall bar particularly at low temperatures.

In low dimensional electron systems, disorder commonly causes localization that manifests as an insulating behavior at low bias, $V$ and low temperature, $T$ when system size is considerably larger than the characteristic localization length, $\xi$. Electron transport in a disordered conductor is governed by variable range hopping (VRH)~\cite{VRH}. In the ohmic regime ($V\ll T$), the conductance is usually described by a stretched exponential behavior with respect to temperature, $G(T) \propto exp[-(T_0/T)^\mu]$, where $\mu$ depends on the dimension of the system. On the other hand, in recent years, a Luttinger-liquid-like behavior is observed in transport experiments on one-dimensional systems made from a wide range of materials~\cite{Loc_exp1,Loc_exp2,Loc_exp3, Loc_exp4, Loc_exp5, Loc_exp6}. The observations are characterized by a conductance that varies in a power law with respect to temperature and bias voltage. However, it was shown by numerical analysis that in quasi-1D systems containing large number of channels, the VRH can also reduce to power law described by the following equations~\cite{PowerLaw_VRH}:

\begin{equation} \label{eq1}
\begin{split}
G \propto T^\alpha ,\ \text{\; \; }\ T\gg (\gamma /2\pi)V,\ \\
G \propto V^\beta ,\ \text{\; \; }\  T\ll (\gamma /2\pi)V,\
\end{split}
\end{equation}
 
\noindent where $\gamma $ is a material and sample dependent parameter that ranged anywhere from 0.01 to 1 in the experiments. Our analysis of temperature-dependence of bulk and edge conductances is in agreement with the aforementioned theory. At sufficiently low temperatures $T\ll (\gamma /2\pi)V$, the edge conductance is nearly independent of temperature whereas at $T\gg (\gamma /2\pi)V$ it scales as power law (Fig.~4). The transition temperatures between the asymptotic saturation and power law behaviors are indicated by dashed line. We conclude a quasi-one-dimensional localization when the Fermi level is tuned inside the normal gap of the system.

\begin{figure}[t]
\begin{center}
\includegraphics[width=0.5\textwidth]{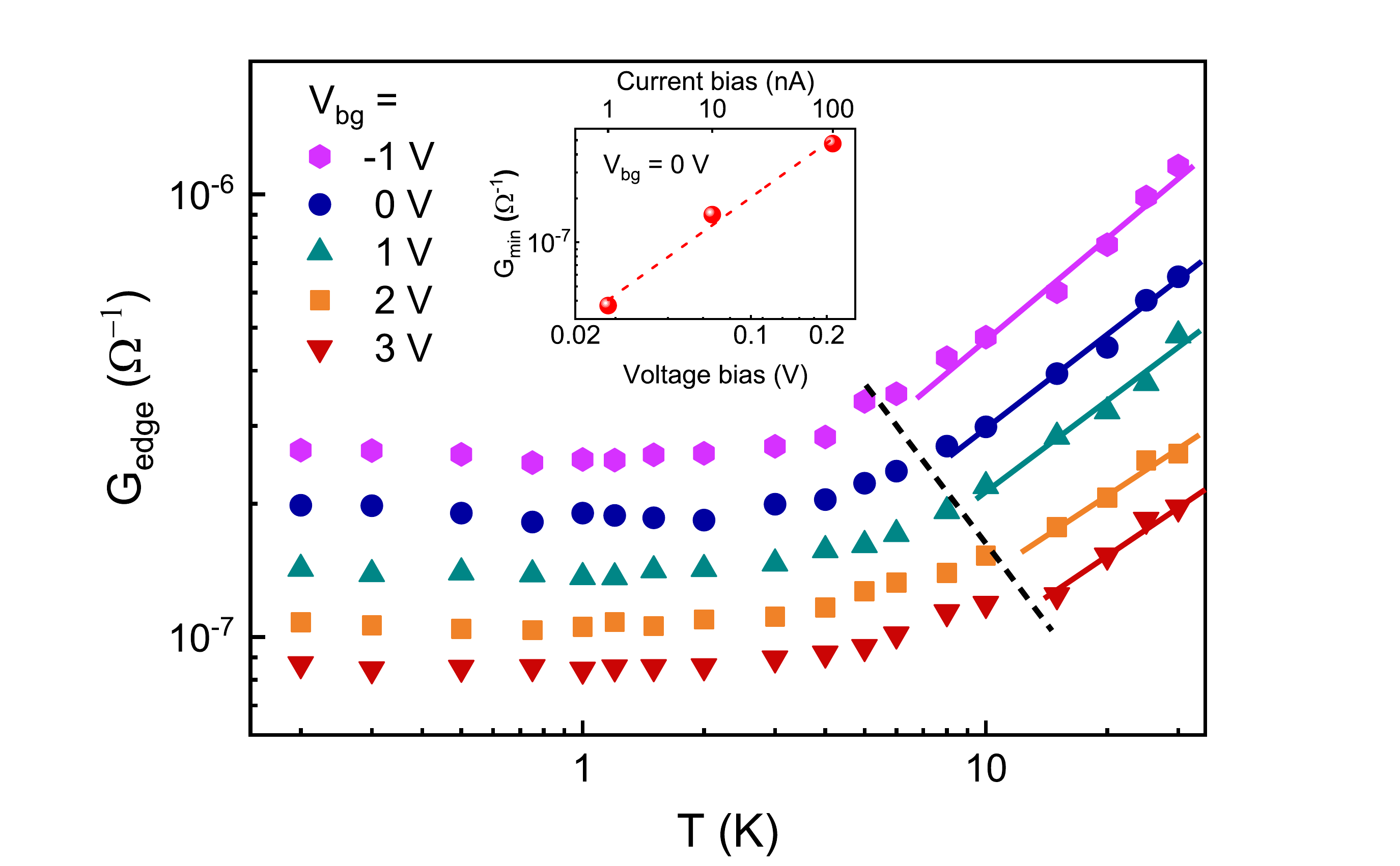}
\caption{\label{Fig4} Edge component of the Hall bar conductance minima in top gate sweeps as a function of temperature. A power law behavior with an average exponent $\alpha = 0.65 \pm 0.05$ at high temperatures (solid lines) and saturated conductance at low temperatures are observed. Dashed line represents $T=(\gamma/2\pi)V$ with $\gamma=0.1$. The inset illustrates the power law scaling of the conductance with voltage excitation at lowest temperature, with a corresponding power exponent $\beta = 1.25 \pm 0.02$. }
\end{center}
\end{figure}

It should be noted that the strong suppression of conductance is also observed in edge lengths as short as $1~\mu$m in a finger gate device in which we studied the length dependency of edge conduction. The dip of longitudinal conductance in top gate sweeps as a function of edge length is shown in Fig. 5. Unlike the existing reports on the quasi-linear length-dependent resistance of the edge channels either in trivial or topological phases, here we have a nonlinear depression of conductance as the length increases in agreement with a localization picture. It is well known that in one dimension, nearly all eigenstates of a disordered system are exponentially localized at low temperatures. Consequently, the conductance decays in an exponential manner as the length increases. The localization length is obtained as $\xi\approx4.5~\mu m$ which leads to a strong localization even in micron-size edges.

\begin{figure}[t]
	\begin{center}
		\includegraphics[width=0.5\textwidth]{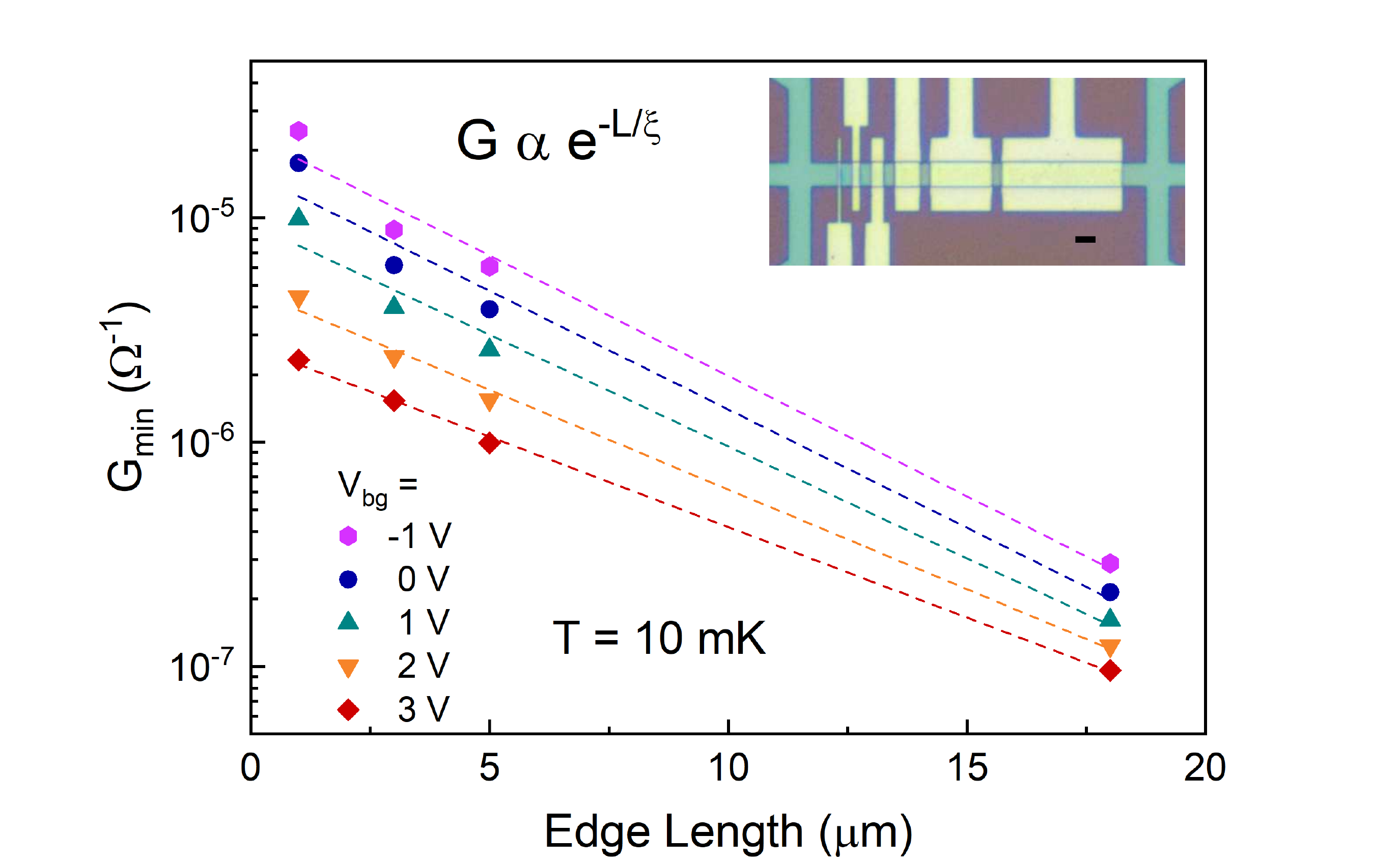}
		\caption{\label{Fig5} The conductance minimum vs edge length measured in the finger gated device.  Optical microscope image is shown in the inset, the black scale bar indicates 10~$\mu m$ length.}\end{center}
\end{figure} 

In conclusion, we observed that in InAs/GaSb bilayer QW structure trivial edge states are strongly localized when InAs layer is  weakly disordered. Our analysis of temperature- and length-dependent edge conduction indicates that silicon atoms that are delta-doped in the InAs layer near the interface of quantum wells are the principal scatterers for e-type carriers as the measured mean free path coincides with the mean spacing between dopants. 2D bulk in the electron band still keeps its high mobility features. The bulk conductivity is characterized as a gapped semiconductor and exponentially diminishes at low temperatures while the conductance due to the edge states prevail in the Hall bar devices. In-depth measurements on Corbino devices and non-local conductivity experiments used to verify the edge states independently. The temperature dependence of the edge conduction in the non-inverted regime is consistent with the quasi-1D localization; scales by power law at high-T and saturates at low-T. The parameter $\gamma$ which defines the transition temperature is determined to be 0.1. Furthermore, at high bias regime, the edge conductance enhances in power law with voltage. These are in general agreement with the previous reports on various 1D systems. The conductance of edge states is measured to scale exponentially with length in the finger gate device and the localization length is calculated as 4.5 $\mu m$. The elimination of undesired trivial edge conductance in InAs/GaSb system is significant towards the realization of a pure QSH phase with only topologically protected helical edge modes. New structures will be explored to overcome the back gate issues and confirm access to the non-trivial phase. 

The authors would like to thank \.{I}nan\c{c} Adagideli for fruitful discussions, and S\"{u}leyman \c{C}elik for his technical support. This research was supported by a Lockheed Martin Corporation Research Grant.
\bibliographystyle{apsrev4-1}
\bibliography{Localization_W1865_arXiv}
\end{document}